\documentclass{sig-alternate}
\usepackage{graphicx}

\usepackage{amssymb,amsfonts,amsmath}

\newcommand{\like}{P} 
\newcommand{\MI}{\textrm{MI}}
\renewcommand{\AA}{\textrm{AA}}
\newcommand{\abs}[1]{\left| #1 \right| }

\DeclareMathOperator*{\argmax}{\textrm{argmax}}
\newcommand{\avg}[1]{\left\langle #1 \right\rangle}
\newcommand{\dP}{\mathrm{d}p}

\begin{document}
\conferenceinfo{KDD'11,} {August 21--24, 2011, San Diego, California, USA.}
\CopyrightYear{2011}
\crdata{978-1-4503-0813-7/11/08}
\clubpenalty=10000
\widowpenalty = 10000

\title{Active Learning for Node Classification\\ in Assortative and Disassortative Networks}
%
%
%
%
%

\numberofauthors{5} 
%
\author{
%
%
\alignauthor
Cristopher Moore\thanks{This work was supported by the McDonnell Foundation.}\\
       \affaddr{Computer Science Dept.} \\
       \affaddr{University of New Mexico Albuquerque NM 87131 USA}\\
       \affaddr{and Santa Fe Institute}\\
       \email{moore@cs.unm.edu}
\alignauthor
Xiaoran Yan\\
       \affaddr{Computer Science Dept.}\\
       \affaddr{University of New Mexico Albuquerque NM 87131 USA}\\
       \email{everyxt@gmail.com}
\alignauthor
Yaojia Zhu\\
       \affaddr{Computer Science Dept.}\\
       \affaddr{University of New Mexico Albuquerque NM 87131 USA}\\
       \email{yaojia.zhu@gmail.com}
\and  
\alignauthor
Jean-Baptiste Rouquier\\
       \affaddr{Complex Systems Institute Rh\^{o}ne-Alpes}\\
       \affaddr{ENS Lyon, France}\\
       \email{jrouquie@gmail.com}
\alignauthor
Terran Lane\\
       \affaddr{Computer Science Dept.}\\
       \affaddr{University of New Mexico Albuquerque NM 87131 USA}\\
       \email{terran@cs.unm.edu}
}

\maketitle
\begin{abstract}
In many real-world networks, nodes have class labels, attributes, or variables that affect the network's topology.  If the topology of the network is known but the labels of the nodes are hidden, we would like to select a small subset of nodes such that, if we knew their labels, we could accurately predict the labels of all the other nodes.  We develop an active learning algorithm for this problem which uses information-theoretic techniques to choose which nodes to explore.  We test our algorithm on networks from three different domains: a social network, a network of English words that appear adjacently in a novel, and a marine food web.  Our algorithm makes no initial assumptions about how the groups connect, and performs well even when faced with quite general types of network structure.  In particular, we do not assume that nodes of the same class are more likely to be connected to each other---only that they connect to the rest of the network in similar ways.
\end{abstract}

\vspace{1mm}
\noindent
{\bf Categories and Subject Descriptors:}
\newline I.2.6 [\textbf{Artificial Intelligence}]: Learning
\newline G.2.2 [\textbf{Discrete Mathematics}]: Graph theory

\vspace{1mm}
\noindent
{\bf General Terms:} Algorithms, Experimentation, Theory  

\vspace{1mm}
\noindent
{\bf Keywords:} complex networks, structure and function, community detection, information theory, active learning,
  collective classification, transductive graph labeling

%
%


\section{Introduction}
In many social, biological, and technological networks, nodes have underlying attributes or variables that are correlated with the network's topology.
Blogs tend to link to other blogs with similar political views~\cite{lada}.  In vertebrate food webs, predators tend to eat prey whose mass is smaller, but not too much smaller, than their own~\cite{brose}.  Networks of word adjacencies are correlated with those words' parts of speech~\cite{newman-leicht}.  In the Internet, different types of service providers form different kinds of links based on their capacities and business relationships~\cite{doyle,rexford}---and so on.

There has been a great deal of work on efficient algorithms for community detection in networks (see~\cite{fortunato,porter} for reviews).  However, most of this work defines a ``community'' as a group of nodes with high density of connections within the group and a low density of connections to the rest of the network.  While this type of \emph{assortative} community structure is common in social networks, we are interested in a more general definition of \emph{functional} community---a group of nodes that connect to the rest of the network in similar ways.  A set of predators might form a functional group in a food web, not because they eat each other, but because they eat similar prey.  In English, nouns often follow adjectives, but seldom follow other nouns.  Even some social networks have \emph{disassortative} structure where pairs of nodes are more likely to be connected if they are from different classes.  For example, some human societies are divided into moieties, and only allow marriages between different moieties~\cite{houseman-white}.

We consider a setting where the topology of the network is known, but the class labels of the nodes are not.  This could be the case, for instance, if we have a network of
blogs and hyperlinks between them (like citations, trackbacks, blogrolls, etc.) and we are trying to classify the blogs according to their political leanings.
Another possible application is in online social networks, where friendships are known and we are trying to infer hidden demographic variables.  This problem is sometimes referred to as collective classification~\cite{sen08}.  However, in that work the focus is on classification of individual nodes.  In contrast, our focus is on the discovery of functional communities in the network, and our underlying generative model is designed around the assumption of that these communities exist.

We make no initial assumptions about the structure of the network---for instance, whether its groups are assortative, disassortative, or some mixture of the two.  We assume that we can learn the label of any given node, but at a cost, say in terms of work in the field or laboratory.  Our goal is to identify a small subset of nodes such that, once we explore them and learn their labels, we can accurately predict the labels of all the others.

We present a general approach to this problem.  Our algorithm uses information-theoretic measures to decide which node to explore next---that is, which one will give us the most information about the rest of the network.
We start with a probabilistic generative model of the network, called a \emph{stochastic block model}~\cite{blockmodel1,blockmodel2}, in which groups connect to each other according to a matrix of probabilities.  This model allows an arbitrary mixture of assortative and disassortative structure, as well as directed links from one group to another, and has been used to model networks in many fields~(e.g. \cite{allesina,hofman,rosvall}).

We stress, however, that our approach could be applied equally well to many other probabilistic models, such as those where nodes belong to a mixture of classes~\cite{airoldi}, a hierarchy of classes and subclasses~\cite{hierarchical}, locations in a latent geographical or social space~\cite{handcock-raftery-tantrum}, or niches in a food web~\cite{niche}.  It could also be applied to degree-corrected block models such as those in~\cite{karrer,morup,degcorrect}, which treat the nodes' degrees as parameters rather than data to be predicted.

At each stage of the learning process, some of the nodes' labels are already known and we need to decide which node to explore next.  We do this by estimating, for each node, the mutual information between its label and the joint distribution of all the others' labels, conditioned on the labels of the nodes that are known so far.  We obtain this estimate by Gibbs sampling, giving each classification of nodes a probability integrated over the parameters of the block model.  We then explore the node for which this mutual information is largest.

A key fact about the mutual information, which we argue is essential to our algorithm's performance, is that it is not just a measure of uncertainty: it is a combination of uncertainty about a node's label and the extent to which it is correlated with the labels of other nodes.  Thus the algorithm explores nodes which maximize the expected amount of information it will gain about the entire network.  It skips nodes whose labels seem obvious to it, or which are uncertain but have little effect on other nodes.  In an assortative network, for instance, it starts by exploring nodes which are central to their communities, and then explores nodes along the boundaries between them, without being told in advance to pursue this strategy.

We also present an alternate approach which maximizes a quantity we call the \emph{average agreement}.  For each node $v$, this is the average number of nodes at which two independent samples of the Gibbs distribution agree, conditioned on the event that they agree at $v$.  Like mutual information, average agreement is high for nodes that are highly correlated with the rest of the network.
A similar idea (but not applied to networks) is present in~\cite{roy_mccallum__error_sampling}.

We test our algorithm on three real-world networks: the social network of a karate club,
a network of common adjacent words in a Charles Dickens novel, and a marine food web of species in the Antarctic.
Each of these networks is curated in the sense that we possess the correct node labels, such as the faction of the social network each individual belongs to, the part of speech of each word, or the part of the habitat each species lives in.  We judge our algorithm according to how accurately it predicts the labels of the unexplored nodes, as a function of the number of nodes it has explored so far.
We also compare our algorithm with several simple heuristics, such as exploring nodes based on their degree or betweenness centrality, and find that it significantly outperforms them.

\section{Related work}

The idea of designing experiments by maximizing the mutual information between the variable we learn next and the joint distribution of the other variables, or equivalently the expected amount of information we gain about the joint distribution, has a long history in statistics, artificial intelligence, and machine learning, e.g.\ Mackay~\cite{mackay} and Guo and Greiner~\cite{mi2}.  Indeed, it goes back to the work of Lindley~\cite{lindley} in the 1950s.  However, to our knowledge this is the first time it has been coupled with a generative model to discover hidden variables in networks.

In recent work, Zhu, Lafferty, and Ghahramani~\cite{zglactive} study active learning of node labels using Gaussian fields and harmonic functions defined using the graph Laplacian.  However, this technique only applies to networks where neighboring nodes are likely to be in the same class---that is, networks with assortative community structure.  In contrast, our techniques are capable of learning about much more general types of network structure, including disassortative and directed relationships between functional communities.

Another approach to active learning of node labels is found in the work of Bilgic and Getoor~\cite{bilgic} and Bilgic, Mihalkova, and Getoor~\cite{bilgicetal}, who use collective vector-based classifiers.  By properly defining the collective relationships between nodes, both assortative or disassortative communities can be learned in this framework. However, our technique differs from theirs by using mutual information as the active learning criterion, which takes into account not just uncertainty, but correlations as well.

Additional works by Goldberg, Zhu, and Wright~\cite{goldberg_zhu_wright} and Tong and Jin~\cite{Tong_Rong__mixed_label_propagation} also perform semi-supervised learning on graphs, and handle the disassortative case.  But they work in a setting where they know, for each link, if the ends should have the same or different labels, such as if one writer quotes another with pejorative words.  In contrast, we work in a setting where we have no such information: only the topology is available to us, and there are no signs on the edges telling us whether we should propagate similar or dissimilar labels.

\section{Model and methods}

We represent our network as a directed graph $G=(V,E)$ with $n$ nodes.  We assume that there are $k$ classes of nodes, so that each node $v$ has a class label $t(v) \in \{1,\ldots,k\}$.  We are given the graph $G$, and our goal is to learn the labels $t(v)$.  To do this, we assume that $G$ is generated by a probabilistic model, in which its topology is correlated with these labels.

The simplest such model, although by no means the only one to which our methods could be applied, is a stochastic block model~\cite{blockmodel1,blockmodel2}.  It assumes that for each pair of nodes $u,v$, there is an edge from $u$ to $v$ with a probability $p_{t(u),t(v)}$ that depends only on their labels, and that these events are independent.  Given a classification, i.e., a function $t:V \to \{1,\ldots,k\}$ assigning a label to each node,
the probability of generating a given graph $G$ in this model is
\begin{align}
\like(G \,|\, t,p)
&= \left( \prod_{(u,v) \in E} p_{t(u),t(v)} \right) \left( \prod_{(u,v) \notin E} (1-p_{t(u),t(v)}) \right)
\nonumber \\
&= \prod_{i,j=1}^k p_{ij}^{e_{ij}} (1-p_{ij})^{n_i n_j - e_{ij}} \, .
\label{eq:like-tp}
\end{align}
Here $n_i = \abs{ \{ v \in V : t(v) = i \} }$ is the number of nodes of class $i$, and $e_{ij} = \abs{ \{ (u,v) \in E : t(u) = i, t(v) = j \} }$ is the number of edges from nodes of class $i$ to nodes of class $j$.  If we wish to focus on undirected graphs, we can modify this expression by restricting the product over pairs of classes with $i \le j$.  We can also forbid self-loops, if we wish, by replacing $n_i^2$ in the term $i=j$ with $n_i (n_i-1)$ or ${n_i \choose 2}$ in the directed or undirected case respectively.

This kind of stochastic block model is well-known in the machine learning, statistics, and network communities~\cite{bickel,jackson,marta,hastings,hofman,rosvall} and has also been used in ecology to identify groups of species in food webs~\cite{allesina}.  Unlike e.g.\ \cite{jackson,hastings,hofman}, we do not assume that $p_{ij}$ takes one value when $i=j$ and a smaller value when $i \ne j$.  In other words, we do not assume an assortative community structure, where nodes are more likely to be connected to other nodes of the same class.  Nor do we require in general that $p_{ij} = p_{ji}$, since the directed nature of the edges may be important---for instance, in a food web or word adjacency network.

If all classifications $t$ are equally likely \emph{a priori}, then Bayes' rule implies that the Gibbs distribution on the classifications, i.e., the probability of $t$ given $G$, is proportional to the probability of $G$ given $t$:
\begin{equation}
\label{eq:gibbs2}
P(t \,|\, G) \propto \like(G \,|\, t) \; .
\end{equation}
In order to define $\like(G \,|\, t)$, we need to integrate $\like(G \,|\, t,p)$ over some prior probability distribution on $p$.  If we assume that the $p_{ij}$ are independent, then this integral factors over the product~\eqref{eq:like-tp}. In particular, if each $p_{ij}$ follows a beta prior, we have the Bayesian estimate of edge probabilities
\begin{align}
\like(G \,|\, t)
&= \iiint_0^1 {\rm d}\{p_{ij}\} \,\like(G \,|\, t,p) \nonumber \\
&= \prod_{i,j=1}^k \int_0^1 \dP_{ij} \,\mathrm{Beta}(p_{ij}|\alpha, \beta) \,p_{ij}^{e_{ij}} (1-p_{ij})^{n_i n_j - e_{ij}} \nonumber \\
&= \prod_{i,j=1}^k \dfrac{ \Gamma (\alpha+\beta)}{\Gamma(\alpha)\Gamma(\beta)}\nonumber \\
&  \int_0^1 \dP_{ij} \,p_{ij}^{e_{ij}+\alpha-1} (1-p_{ij})^{n_i n_j - e_{ij}+\beta-1} \nonumber \\
&= \prod_{i,j=1}^k \dfrac{ \Gamma (\alpha+\beta)}{\Gamma(\alpha) \,\Gamma(\beta)} \dfrac{\Gamma(e_{ij}+\alpha) \,\Gamma(n_i n_j-e_{ij}+\beta)}{ \Gamma (n_i n_j+\alpha+\beta)} \; .
\end{align}
For reasonable choices of the hyperparameters $\alpha$ and $\beta$, the prior dominates only in small data cases, such as very small networks or sparsely populated classes.  For such small data cases, the beta prior allows the user to input some domain knowledge about, say, the (dis)assortativity of the target network's community structure.  In the limit of large data, the prior will wash out and the data-driven community structure will dominate.

If the user wishes to remain agnostic, however, he or she can specify a uniform prior ($\alpha=\beta=1$) and allow the learning algorithm to estimate the degree of assortativity, disassortativity, directedness, and so on entirely from the data. We take this approach in this paper, in which case
\begin{align}
\like(G \,|\, t)
&= \prod_{i,j=1}^k \frac{1}{(n_i n_j + 1) {n_i n_j \choose e_{ij}}} \; . \label{eq:gibbs}
\end{align}

An even simpler approach is to assume that the $p_{ij}$ take their maximum likelihood values
\begin{equation}
\hat{p}_{ij} = \argmax_p \,\like(G \,|\, t,p) = e_{ij} / n_i n_j \; ,
\end{equation}
and set $\like(G \,|\, t) = \like(G \,|\, t,\hat{p})$.  This approach was used, for instance, for a hierarchical block model in~\cite{hierarchical}.  When $k$ is fixed and the $n_i$ are large, this will give results similar to~\eqref{eq:gibbs}, since the integral over $p$ is tightly peaked around $\hat{p}$.  However, for any particular finite graph it makes more sense, at least to a Bayesian, to integrate over the $p_{ij}$, since they obey a posterior distribution rather than taking a fixed value.  Moreover, averaging over the parameters as in~\eqref{eq:gibbs} discourages overfitting, since the average likelihood goes down when we increase $k$ and hence the volume of the parameter space.  This gives us a principled way to determine $k$ automatically, although in this paper we set $k$ by hand.  Other methods to determine $k$ include minimum description length (MDL) techniques~\cite{rosvall} and the Akaike information criterion~\cite{allesina}

\section{Active Learning}

In the active learning setting, the algorithm can learn the class label of any given node, but at a cost---say, by devoting resources in the laboratory or the field.  Since these resources are limited, it has to decide which node to explore.  Its goal is to explore a small set of nodes and use their labels to guess the labels of the remaining nodes.

One natural approach is to explore the node $v$ with the largest mutual information (MI) between its label $t(v)$ and the labels $t(G \setminus v)$ of the other nodes according to the Gibbs distribution~\eqref{eq:gibbs2}.  We can write this as the difference between the entropy of $t(G \setminus v)$ and its conditional entropy given $t(v)$,
\begin{equation}
\MI(v) = I(v ; G \setminus v) = H(G \setminus v) - H(G \setminus v \,|\, v) \; .
\end{equation}
Here $H(G \setminus v \,|\, v)$ is the entropy, averaged over $t(v)$ according to the marginal of $t(v)$ in the Gibbs distribution, of the joint distribution of $t(G \setminus v)$ conditioned on $t(v)$.  In other words, $\MI(v)$ is the expected amount of information we will gain about $t(G \setminus v)$, or equivalently the expected decrease in the entropy, that will result from learning $t(v)$.

Since the mutual information is symmetric, we also have
\begin{equation}
\MI(v) = I(v ; G \setminus v) = H(v) - H(v \,|\, G \setminus v) \; ,
\end{equation}
where $H(v)$ is the entropy of the marginal distribution of $t(v)$, and $H( v \,|\, G \setminus v)$ is the entropy, on average, of the distribution of $t(v)$ conditioned on the labels of the other nodes.  Thus $\MI(v)$ is large if (i) we are uncertain about $v$, so that $H(v)$ is large, and (ii) $v$ is strongly correlated with the other nodes, so that $H(v \,|\, G \setminus v)$ is small.

We estimate these entropies by sampling from the space of classifications $t$ according to the Gibbs distribution.
Specifically, we use a single-site heat-bath Markov chain.  At each step, it chooses a node $v$ uniformly from among the unexplored nodes, and chooses its label $t(v)$ according to the conditional distribution proportional to $\like(G \,|\, t)$, assuming that the labels of all other nodes stay fixed.  In addition to exploring the space, this allows us to collect a sample of the conditional distribution of the chosen node $v$ and its entropy.  Since $H( v \,|\, G \setminus v)$ is the average of the conditional entropy, and since $H(v)$ is the entropy of the average conditional distribution, we can write
\begin{equation}
I(v ; G \setminus v) = - \sum_{i=1}^k \avg{P_i} \ln \avg{P_i} + \avg{ \sum_{i=1}^k P_i \ln P_i} \; ,
\end{equation}
where $P_i$ is the probability that $t(v)=i$ and $\avg{\cdot}$ denotes the average, according to the Gibbs distribution, over the labels of the other nodes.

We offer no theoretical guarantees about the mixing time of this Markov chain, and it is easy to see that there are families of graphs and values of $k$ for which it it takes exponential time.
However, for the real-world networks we have tried so far, it appears to converge to equilibrium
in a reasonable amount of time.  We test for equilibrium by measuring whether the marginals change noticeably when the number of updates is increased by a factor of $2$.  We improve our estimates by averaging over many runs, each one starting from an independently random initial state.

We say that the algorithm is in \emph{stage $j$} if it has already explored $j$ nodes.  In that stage, it estimates $\MI(v)$ for each unexplored node $v$, using the Markov chain to sample from the Gibbs distribution conditioned on the labels of the nodes explored so far.  It then explores the node $v$ with the largest MI.  We provide it with the correct value of $t(v)$ from the curated network, and it moves on to the next stage.

The mutual information is not the only quantity we might use to identify which node to explore.  Another is the \emph{average agreement}, which we define as follows.  Given two classifications $t_1, t_2$, define their \emph{agreement} as the number of nodes on whose labels they agree,
\begin{equation}
\abs{t_1 \cap t_2} = \abs{ \{ v : t_1(v) = t_2(v) \} } \; .
\end{equation}
Since our goal is to label as many nodes correctly as possible, we wish we could maximize the agreement between an classification $t_1$, drawn from the Gibbs distribution, and the correct classification $t_2$.  However, the algorithm doesn't know $t_2$, so it assumes that it is drawn from the Gibbs distribution as well.  Exploring $v$ projects onto the part of the joint distribution of $(t_1,t_2)$ where $t_1(v)=t_2(v)$.  So, we define $\AA(v)$ as the expected agreement between two classifications $t_1, t_2$ drawn independently from the Gibbs distribution, conditioned on the event that they agree at $v$:
\begin{equation}
\label{eq:aa}
\AA(v) = \frac{\sum_{t_1,t_2: t_1(v)=t_2(v)} P(t_1) P(t_2) \abs{ t_1 \cap t_2 }}{\sum_{t_1,t_2: t_1(v)=t_2(v)} P(t_1) P(t_2)} \, .
\end{equation}
We estimate the numerator and denominator of $\AA(v)$ using the same heat-bath Gibbs sampler as for $\MI(v)$, except that we sample independent pairs of classifications $(t_1,t_2)$ by starting the Markov chain at two independently random initial states.


\section{Results and discussion}

\begin{figure}
\centering
\includegraphics[width=0.49\textwidth]{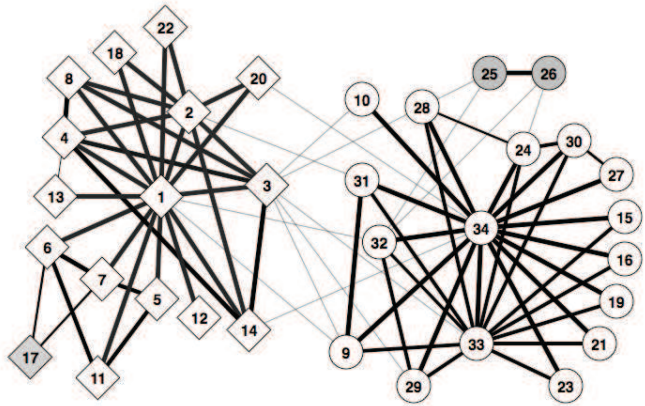}
\caption{Zachary's Karate Club.}
\label{fig:karate}
\end{figure}

We tested our algorithms on three different networks from three different fields.  The first is Zachary's Karate Club~\cite{zachary}.  As shown in Fig.~\ref{fig:karate}, this is a social network consisting of $34$ members of a karate club, where undirected edges represent friendships.  The club split into two factions, indicated by diamonds and circles respectively. One of them centered around the instructor (node 1) and the other around the club president (node 34), each of which formed their own club. Shaded nodes are more peripheral, and have weaker ties to their communities.  This network is highly assortative, with a high density of edges within each faction and a low density of edges between them.

\begin{figure}
\centering
\epsfig{file=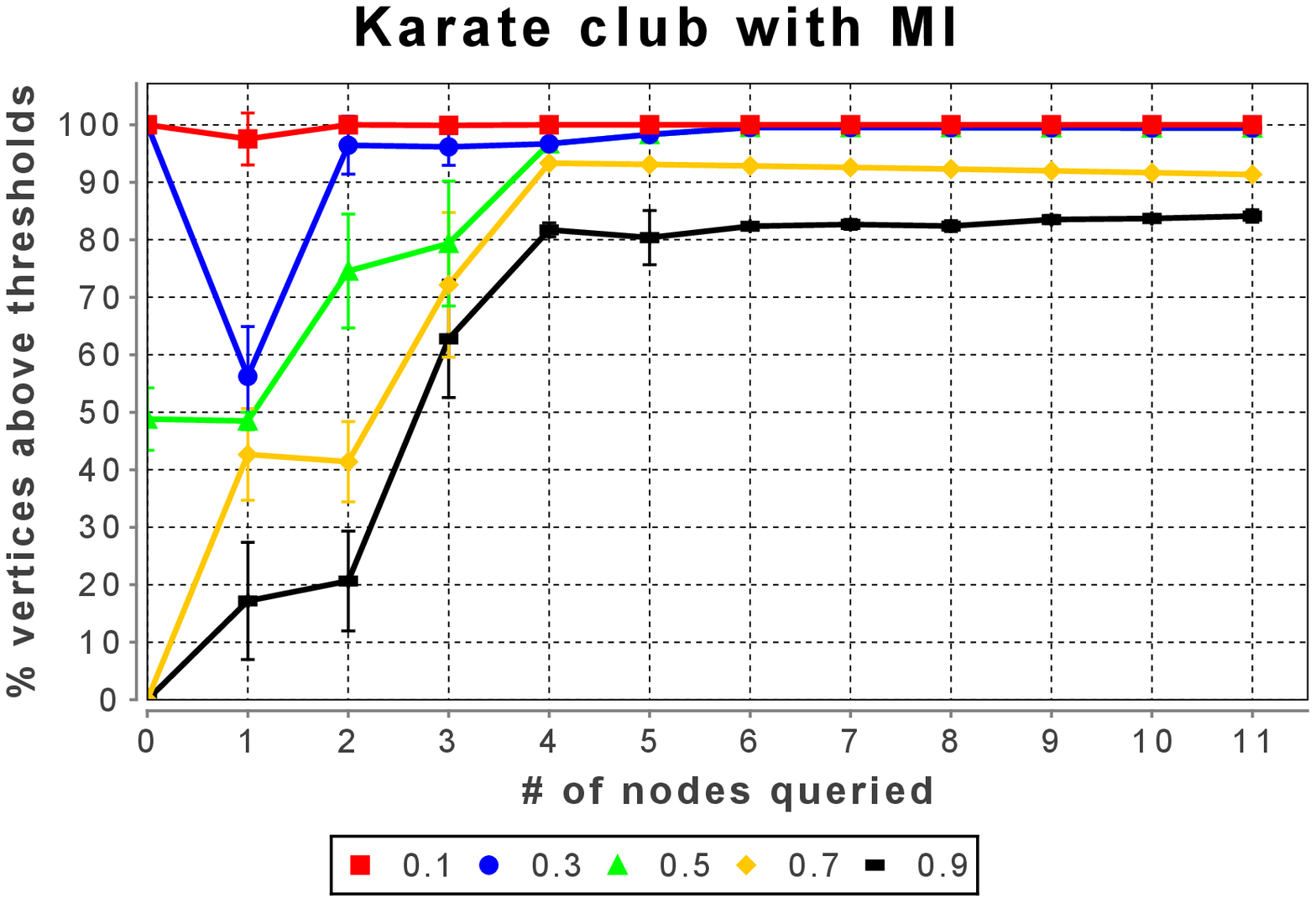, height=2in, width=3in}
\epsfig{file=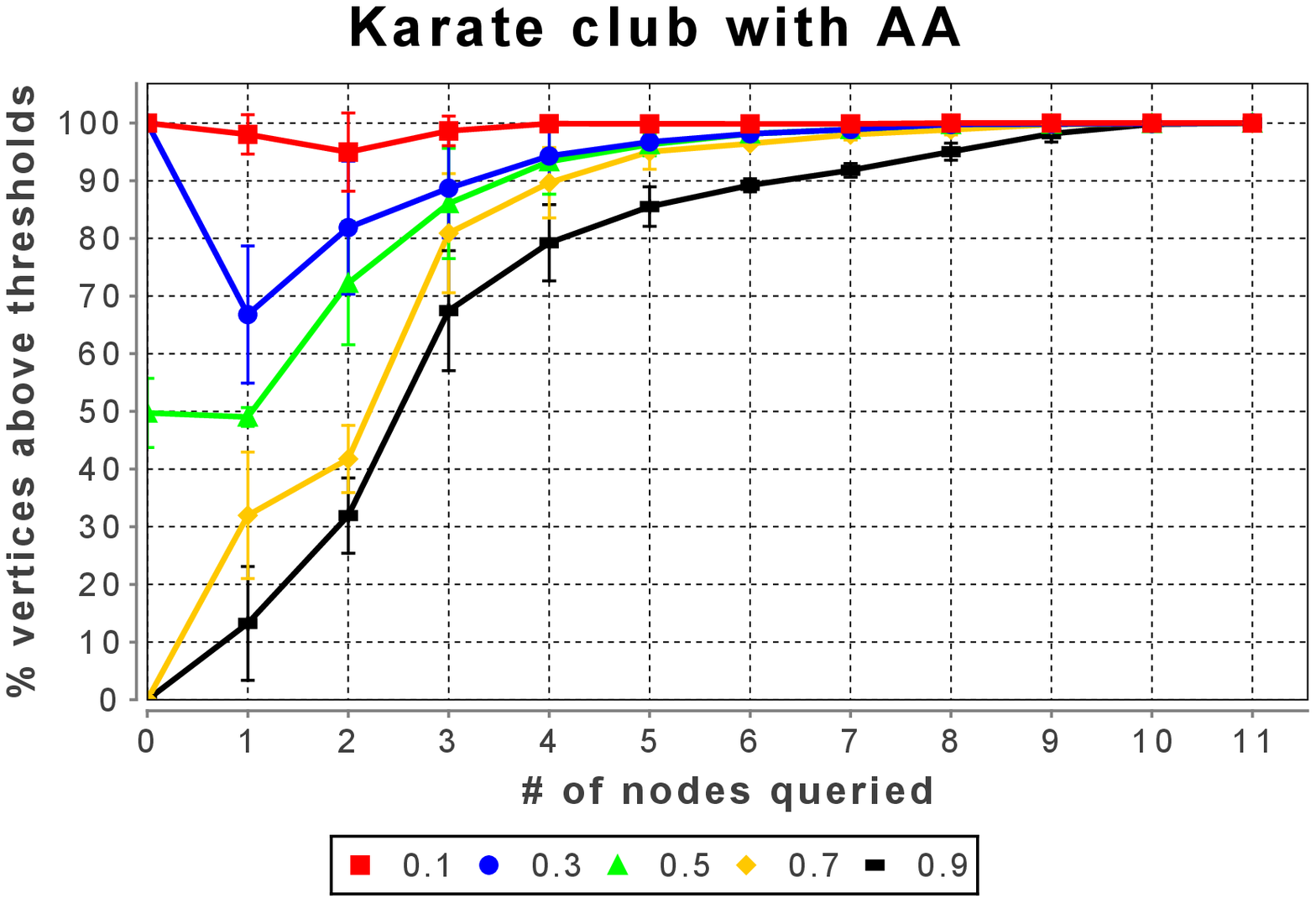, height=2in, width=3in}
\caption{Results of the active learning algorithms on Zachary's Karate Club network.}
\label{fig:learnkarate}
\end{figure}


We judge the performance of each algorithm by asking, at each stage and for each node, with what probability the Gibbs distribution assigns it the correct label.  In each stage we sampled the Gibbs distribution using $100$ independently chosen initial conditions, doing $2 \times 10^4$ steps of the heat-bath Markov chain for each one, and computing averages using the last $10^4$ steps.  Increasing the number of Markov chain steps to $10^5$ per stage produced only marginal improvements in performance. Fig.~\ref{fig:learnkarate} shows what fraction of the unexplored nodes are assigned the correct label with probability at least $q$, for various thresholds $q=0.1, 0.3, 0.5, 0.7, 0.9$, as a function of the stage $j$.

After exploring just four or five nodes, both of our algorithms succeed in correctly predicting the labels of most of the remaining nodes---i.e., to which faction they belong---with high accuracy.  The AA algorithm performs slightly better than MI, achieving an accuracy close to $100\%$ after exploring nine nodes.  Of course, the Karate Club network is quite small, and there are many community-finding algorithms that classify the two factions with perfect or near-perfect accuracy~\cite{porter,fortunato}.

\begin{figure}
\centering
\epsfig{file=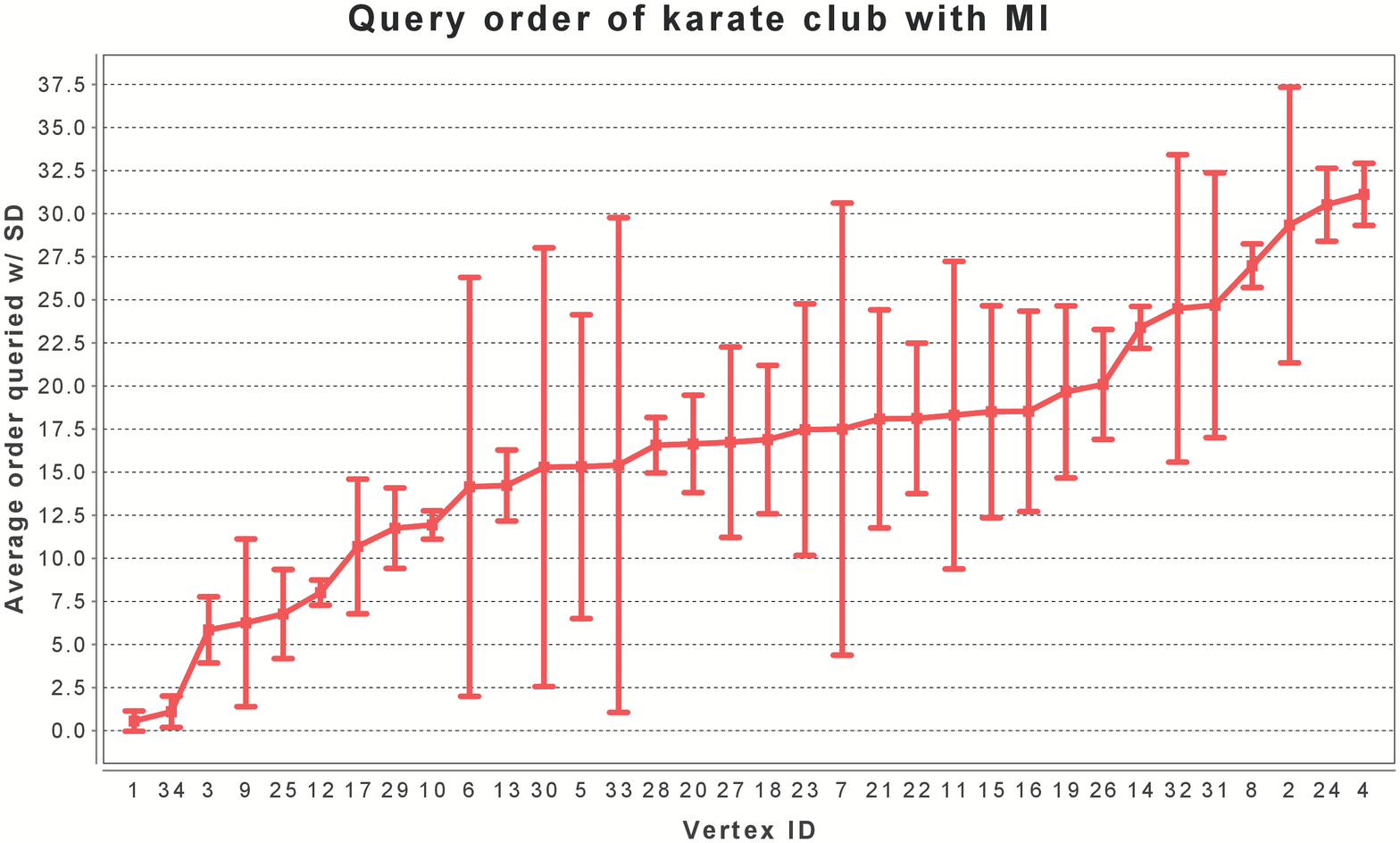, height=2in, width=3in}
\epsfig{file=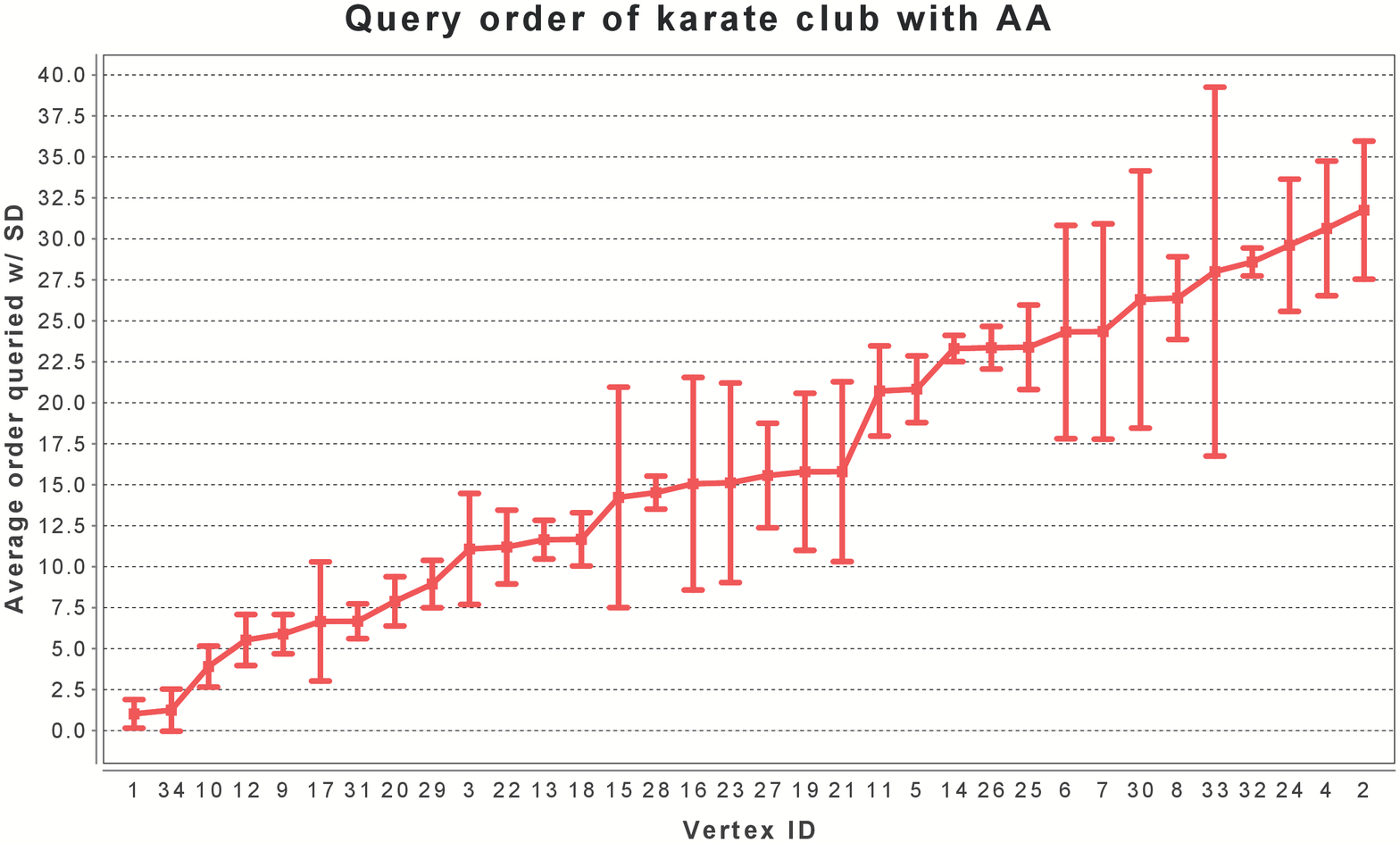, height=2in, width=3in}
\caption{The order in which the active learning algorithms explore nodes in Zachary's Karate Club.}
\label{fig:explorekarate}
\end{figure}

Perhaps more interesting is the \emph{order} in which our algorithms choose to explore the nodes.  In Fig.~\ref{fig:explorekarate}, we sort the nodes in order of the median stage at which they are explored.  Error bars show $90\%$ confidence intervals over $100$ independent runs of each algorithm.  Some nodes show a large variance in the stage in which they are explored, while others are consistently explored at the beginning or end of the process.  Both algorithms start by exploring nodes $1$ and $34$, which are central to their respective communities.  Note that these nodes are chosen, as we argued above, not just because their labels are uncertain, but because they are highly correlated with the labels of other nodes.

After learning that nodes $1$ and $34$ are in class $1$ and $2$ respectively, the algorithms ``know'' that the network consists of two assortative communities.  They they explore nodes such as 3, 9, and 10 which lie at the boundary between these communities.  Once the boundary is clear, they can easily predict the labels of the remaining nodes.  The last nodes to be explored are those such as $2$, $4$, and $24$, which lie so deep inside their communities that their  labels are not in doubt.


The second network consists of the 60 most commonly occurring nouns and the 60 most commonly occurring adjectives in Charles Dickens' novel \emph{David Copperfield}.  A directed edge connects any pair of words that appear adjacently in the text, pointing from the preceding word to the following one.  Excluding eight words which are disconnected from the rest leaves a network with 112 nodes~\cite{newman_eigen}.  Unlike Zachary's Karate Club, this network is both directed and highly disassortative.  Of the 1494 edges, 1123 of them point from adjectives to nouns.  This lets us classify most nodes early on, simply by labeling a node as an adjective or noun if its out-degree or in-degree is large.

\begin{figure}
\centering
\includegraphics[width=0.49\textwidth]{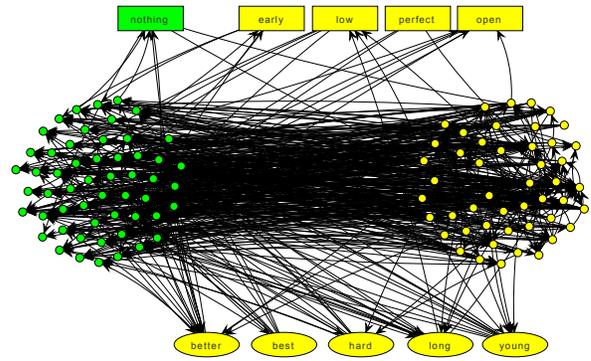}
\caption{The order in which the active learning algorithm MI explores nodes in word adjacency network from the novel \textit{David Copperfield}.}
\label{fig:exploreWords}
\end{figure}

\begin{figure*}
\centering
\includegraphics[width=0.46\textwidth]{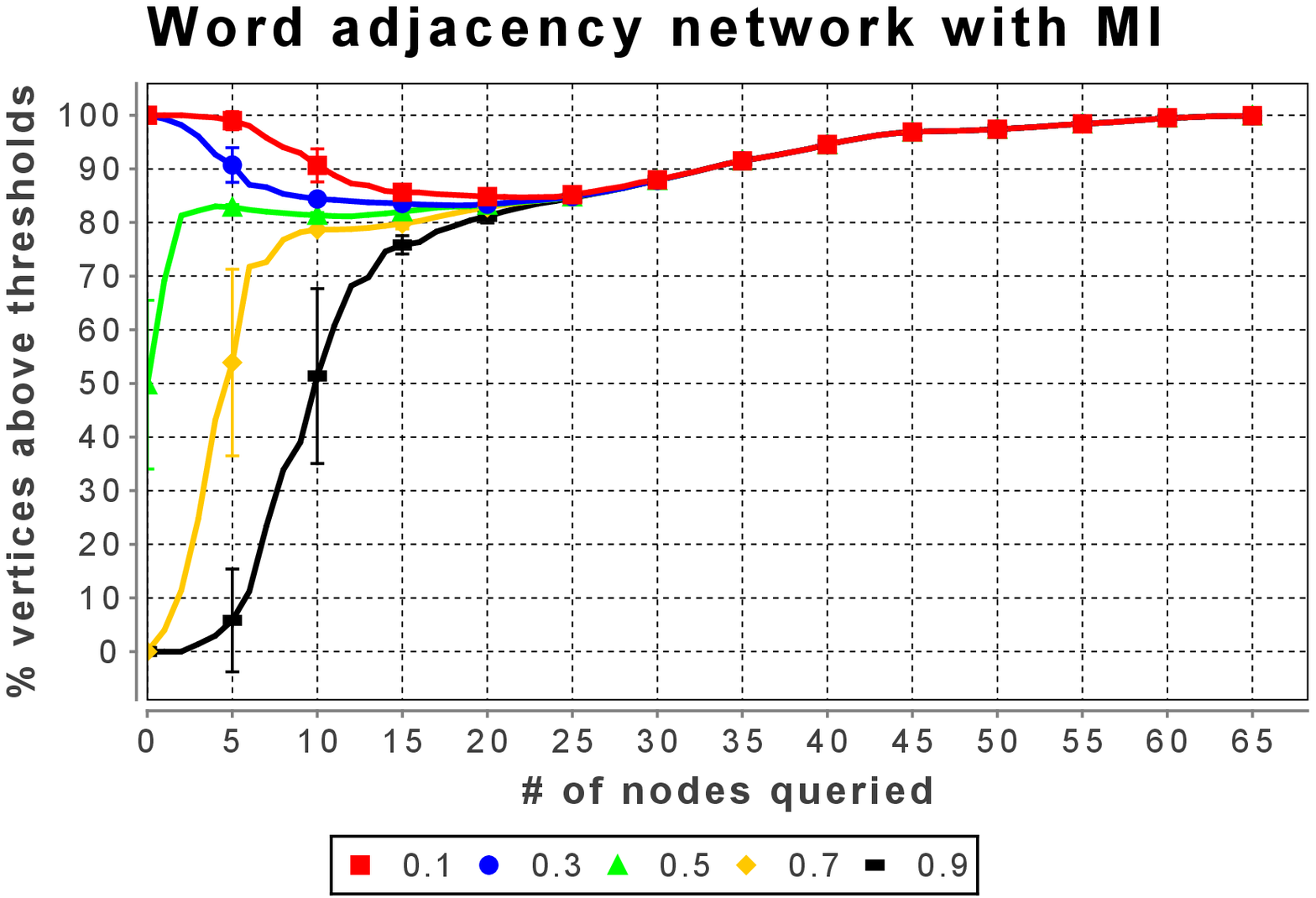}
\includegraphics[width=0.46\textwidth]{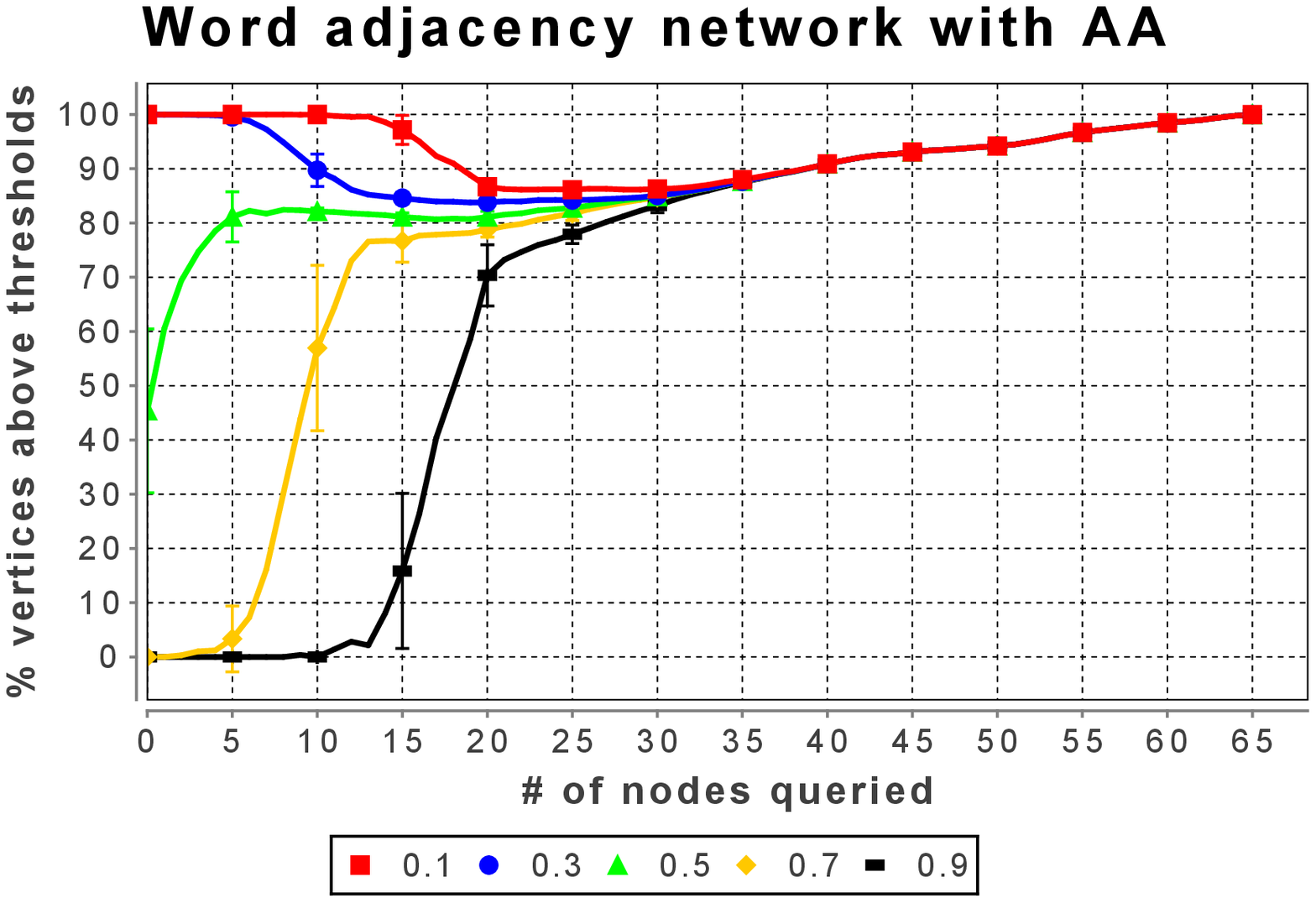}
\caption{Results of the active learning algorithms on word adjacency network in the novel \textit{David Copperfield} by Charles Dickens.}
\label{fig:learnWord}
\end{figure*}

Accordingly, our algorithms focus their attention on words about which they are uncertain, like ``early,'' ``low,'' and ``nothing,'' whose out-degrees and in-degrees  in the text are roughly equal, and words like ``perfect'' that precede words of both classes (see Fig.~\ref{fig:exploreWords}, where green and yellow nodes represent nouns and adjectives respectively; rectangular nodes are explored first, and elliptical ones last). Once these nodes are resolved, both algorithms achieve high accuracy---$80\%$ accuracy after exploring 20 nodes and close to $100\%$ after exploring 65 nodes (see Fig.~\ref{fig:learnWord}).

\begin{figure*}
\centering
\includegraphics[width=0.46\textwidth]{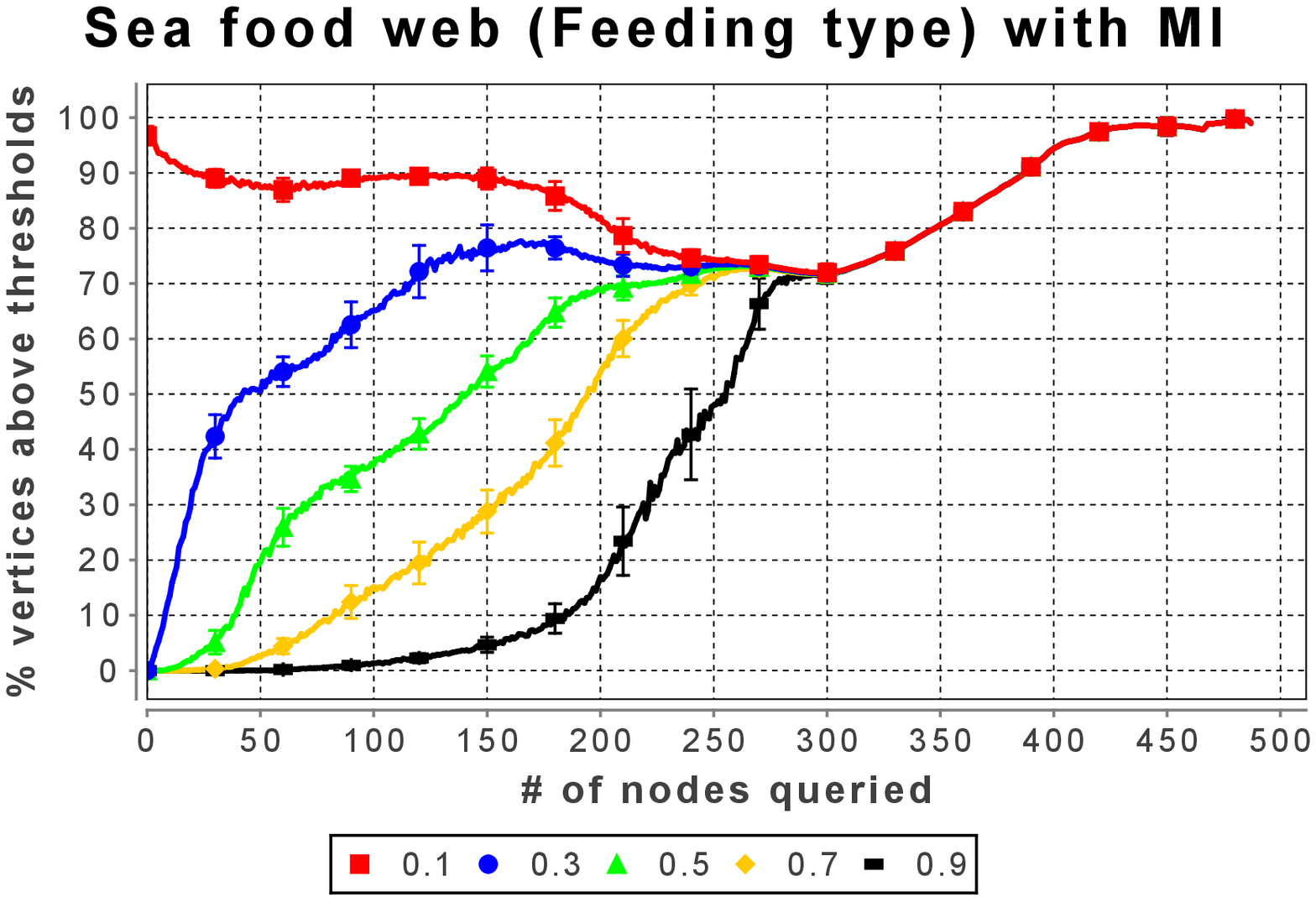}
\includegraphics[width=0.46\textwidth]{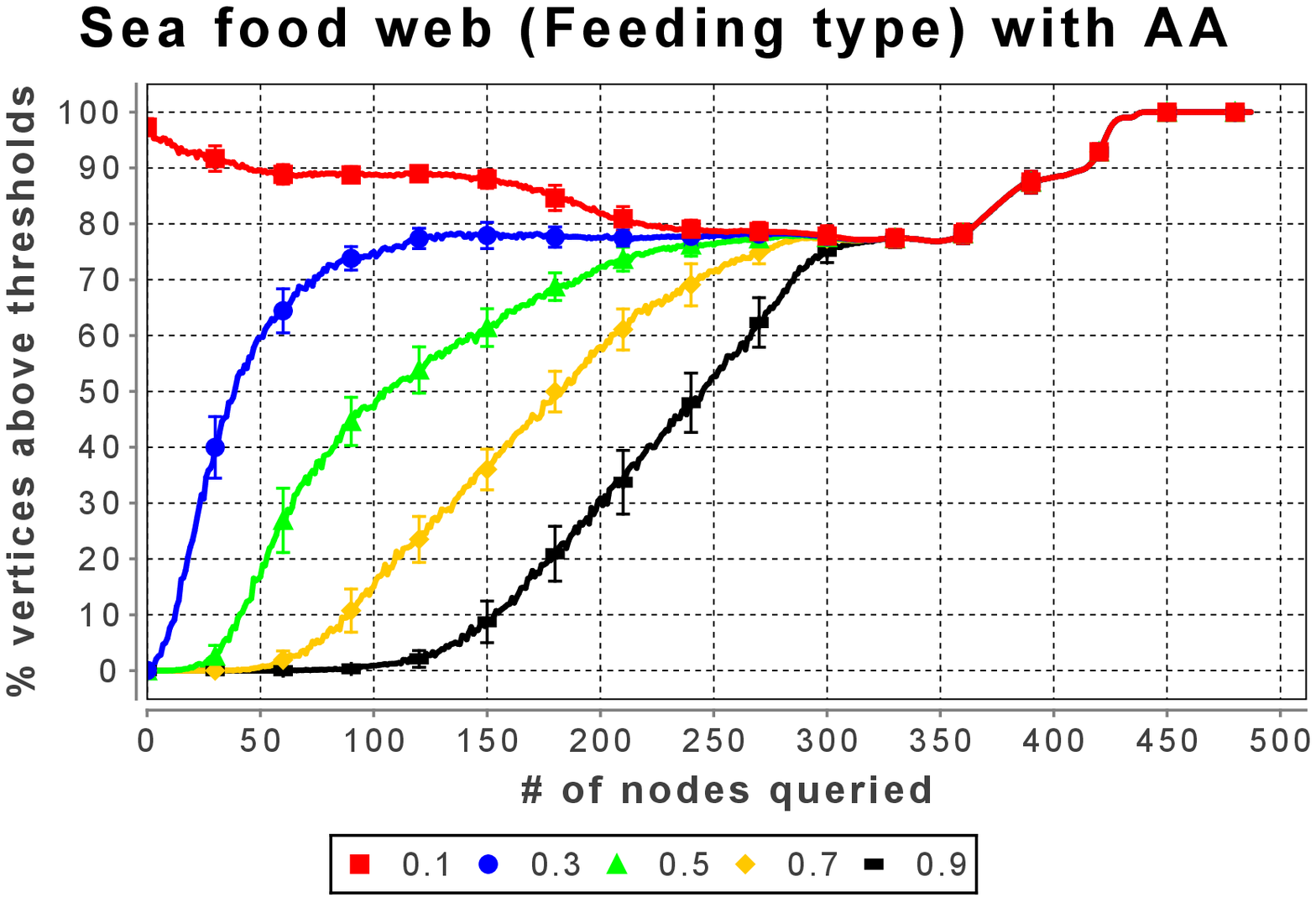}\\
\includegraphics[width=0.46\textwidth]{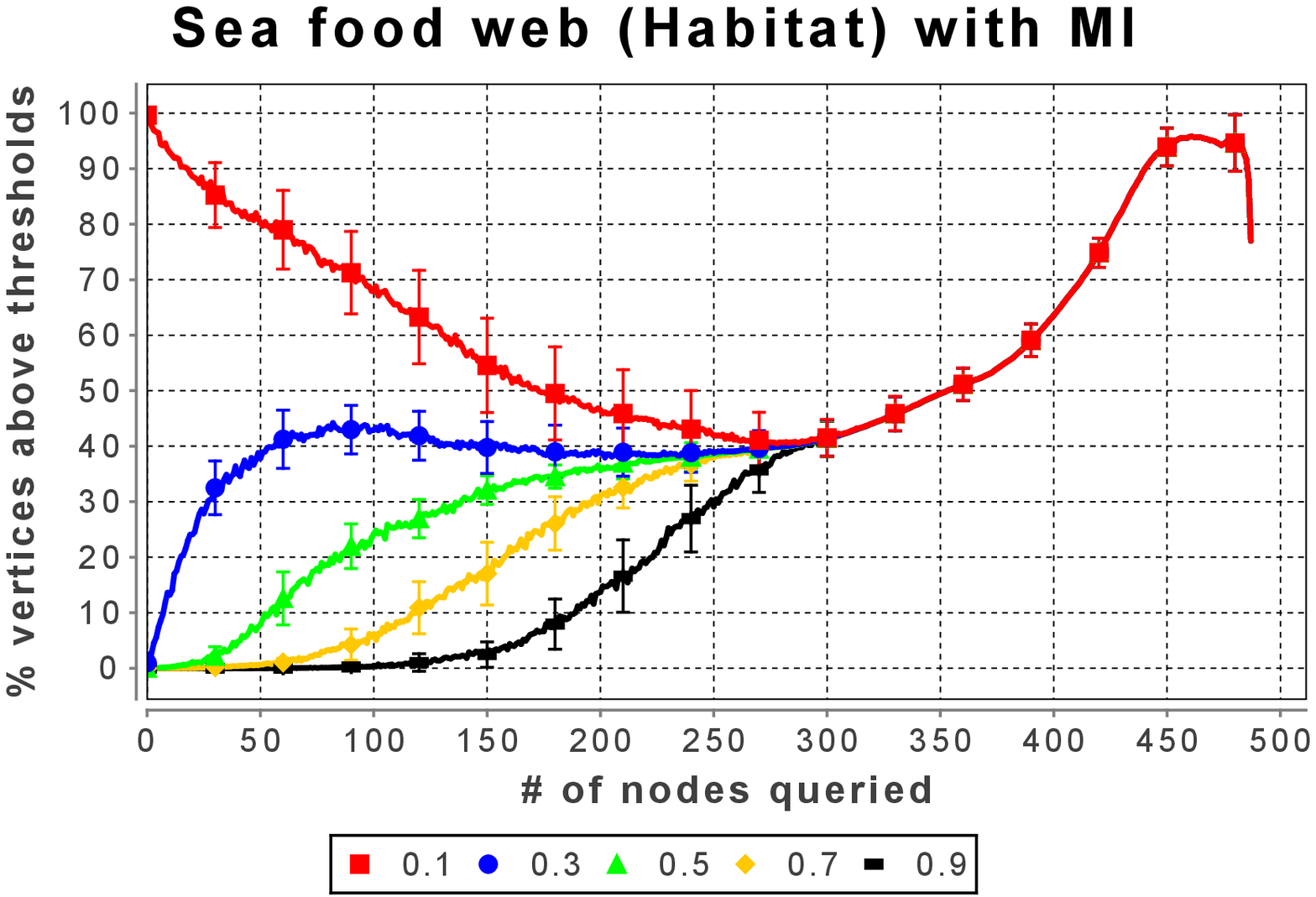}
\includegraphics[width=0.46\textwidth]{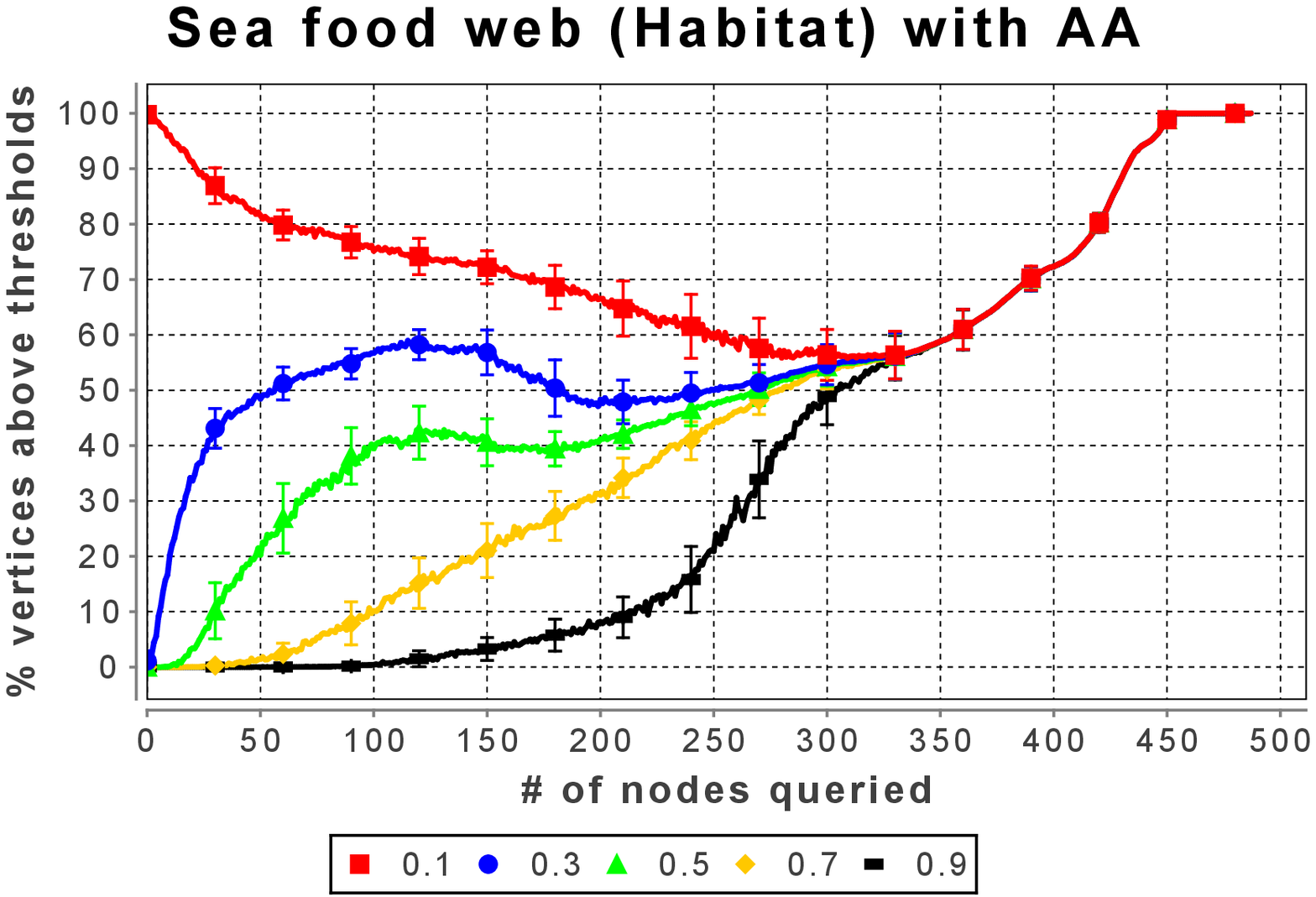}
\caption{Results for the Weddell Sea food web.}
\label{fig:learnfoodweb}
\end{figure*}


In each stage we sampled the Gibbs distribution using $100$ independently chosen initial conditions, doing $5\times 10^4$ steps of the heat-bath Markov chain for each one, and computing averages using the last $2.5 \times 10^4$ steps. Increasing the number of Markov chain steps to $10^5$ per stage produced only marginal improvements in performance.  As in Fig.~\ref{fig:learnkarate}, the $y$-axis shows the fraction of unexplored nodes which are labeled correctly by the conditional Gibbs distribution with probability at least $q$, for $q=0.1, 0.3, 0.5, 0.7, 0.9$. The performance of the two algorithms is similar in the later stages, but unlike the Karate Club, here MI performs noticeably better than AA in the early stages.


The third network is a food web of $488$ species in the Weddell Sea in the Antarctic~\cite{brose,ute2,ute}, with edges pointing to each predator from its prey.  This data set is very rich, but we focus on two particular variables---the feeding type and the habitat in which the species lives. The feeding type takes $k=6$ values, namely primary producer, omnivorous, herbivorous/detrivorous, carnivorous, detrivorous, and carnivorous/necrovorous. The habitat variable takes $k=5$ values, namely pelagic, benthic, benthopelagic, demersal, and land-based.

We show results of our algorithms for both variables in Fig.~\ref{fig:learnfoodweb}.  The results are averaged over 100 runs of each algorithm.  In each stage we sampled the Gibbs distribution using $100$ independently chosen initial conditions, doing $5\times 10^4$ steps of the heat-bath Markov chain for each one, and computing averages using the last $2.5 \times 10^4$ steps. For the feeding type, after exploring half the nodes, both algorithms correctly label about $75\%$ of the remaining nodes.  For the habitat variable, both algorithms are less accurate, although AA performs somewhat better than MI.  Note that the accuracy only includes the unexplored nodes, not the nodes we have already explored.  Thus it can decrease if we explore easily-classified nodes early on, so that hard-to-classify nodes form a larger fraction of the remaining ones.

Fig.~\ref{fig:learnfoodweb} shows that both algorithms get to a state where they are confident, but wrong, about many of the unexplored nodes.  For the feeding type variable, for instance, after the AA algorithm has explored $300$ species, it labels $75\%$ of the remaining nodes correctly with probability $90\%$, but it labels the other $25\%$ correctly with probability less than $10\%$.  In other words, it has a high degree of confidence about all the nodes, but is wrong about many of them.  Its accuracy improves as it explores more nodes, but it doesn't achieve high accuracy on all the unexplored nodes until there are only about $60$ of them left.

Why is this?  We argue that the fault lies, not with our learning algorithms and the order in which they explore the nodes, but with the stochastic block model and its ability to model the data.  For example, for the habitat variable, these algorithms perform well on pelagic, demersal, and land-based species.  But the benthic habitat, which is the largest and most diverse, includes species with many feeding types and trophic levels.

These additional variables have a large effect on the topology, but they are not taken into account by the block model.
As a result, more than half the benthic species are mislabeled by the block model in the following sense: even if we condition on the correct habitats of \emph{all} the other species, the species' most likely habitat is pelagic, benthopelagic, demersal, or land-based.  Specifically, 219 of the 488 species are mislabeled by the most likely block model, $94\%$ of them with confidence over $0.9$.

Of course, we can also regard our algorithms' mistakes as evidence that these habitat classifications are not cut and dried.  Indeed, ecologists recognize that there are ``connector species'' that connect one habitat to another, and belong to some extent to both.

To test our hypothesis that it is the block model's inability to model the data that causes some nodes to be misclassified, we artificially modified the data set to make it consistent with the block model.  Starting with the nodes' original class labels, we updated the habitat of each species to its most likely value according to the block model, given the habitats of all the other species.  After iterating this process six times,
we reached a fixed point where each species' habitat is consistent with the block model's predictions.  On this synthetic data set both of our learning algorithms perform perfectly, predicting the habitat of every species with close to $100\%$ accuracy after exploring just $18\%$ of them.

More generally, it is important to remember that the topology of the network is only imperfectly correlated with the nodes' types.  Zachary~\cite{zachary} relates that one of members of the Karate Club joined the instructor's faction even though the network's topology suggests that he was more strongly connected to the president.  The reason is that he was only three weeks away from a test for his black belt when the split occurred.  He had already invested four years learning the instructor's style of karate, and if he had joined the president's club he would have had to start over with a white belt.  In any real-world network, there is information of this kind that is not reflected in the topology and which is hidden from our algorithm.  If a node is of a given class for idiosyncratic reasons like these, we cannot expect any algorithm based solely on topology and the other nodes' class labels---no matter how sophisticated a probabilistic model we use---to correctly classify it.

\section{Comparison with Simple\\ Heuristics}

\begin{figure*}
\centering
\includegraphics[width=0.46\textwidth]{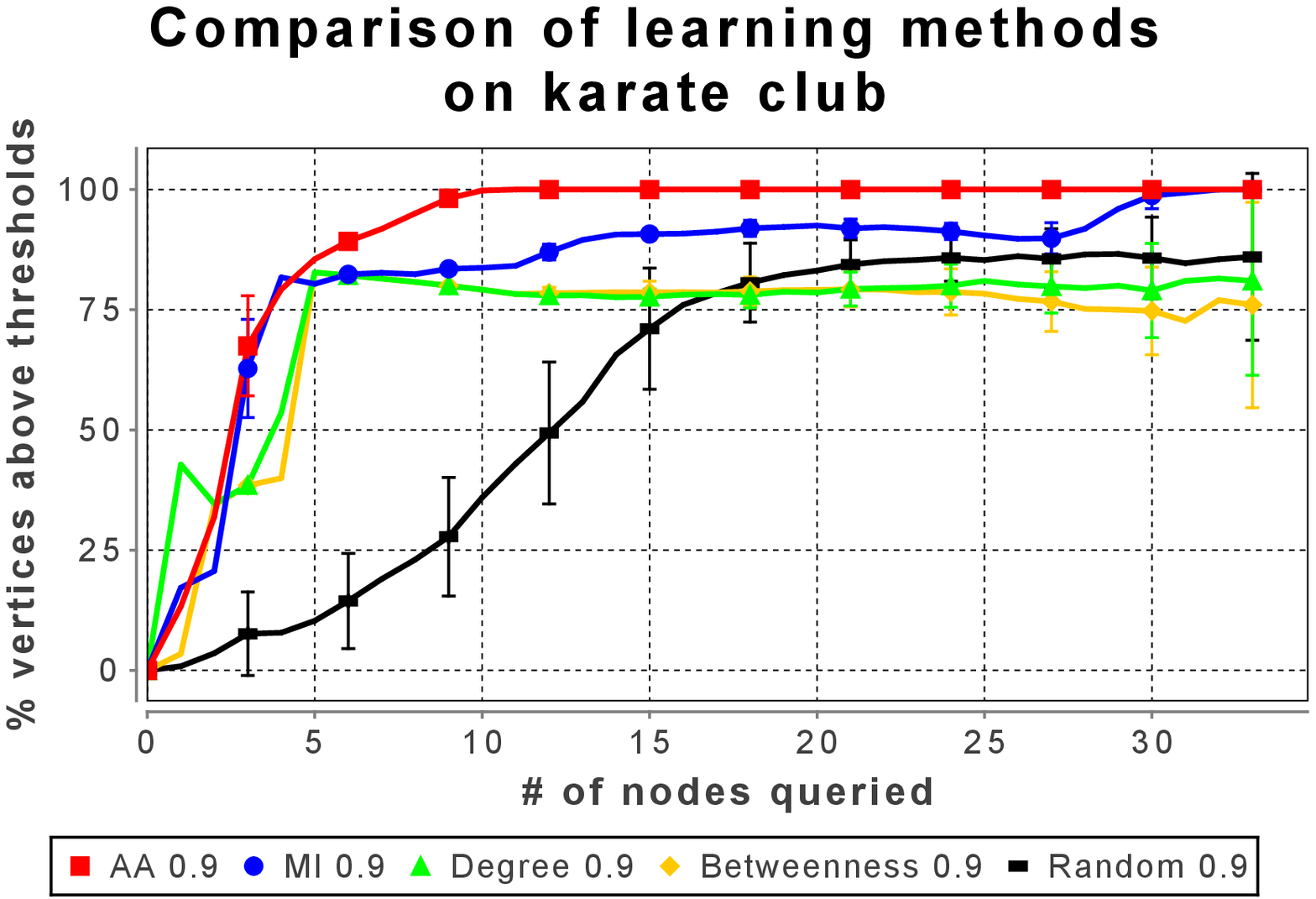}
\includegraphics[width=0.46\textwidth]{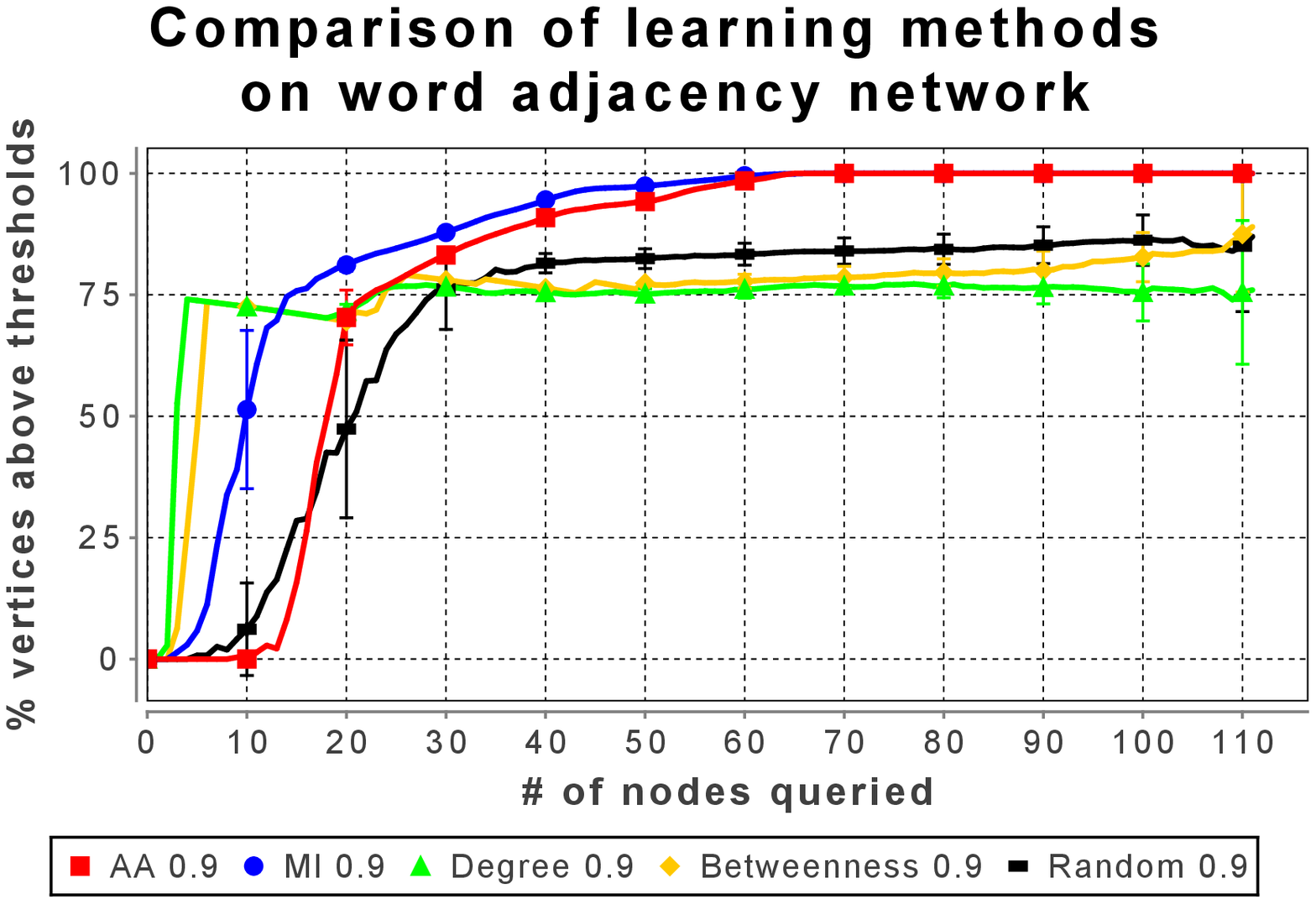}\\
\includegraphics[width=0.46\textwidth]{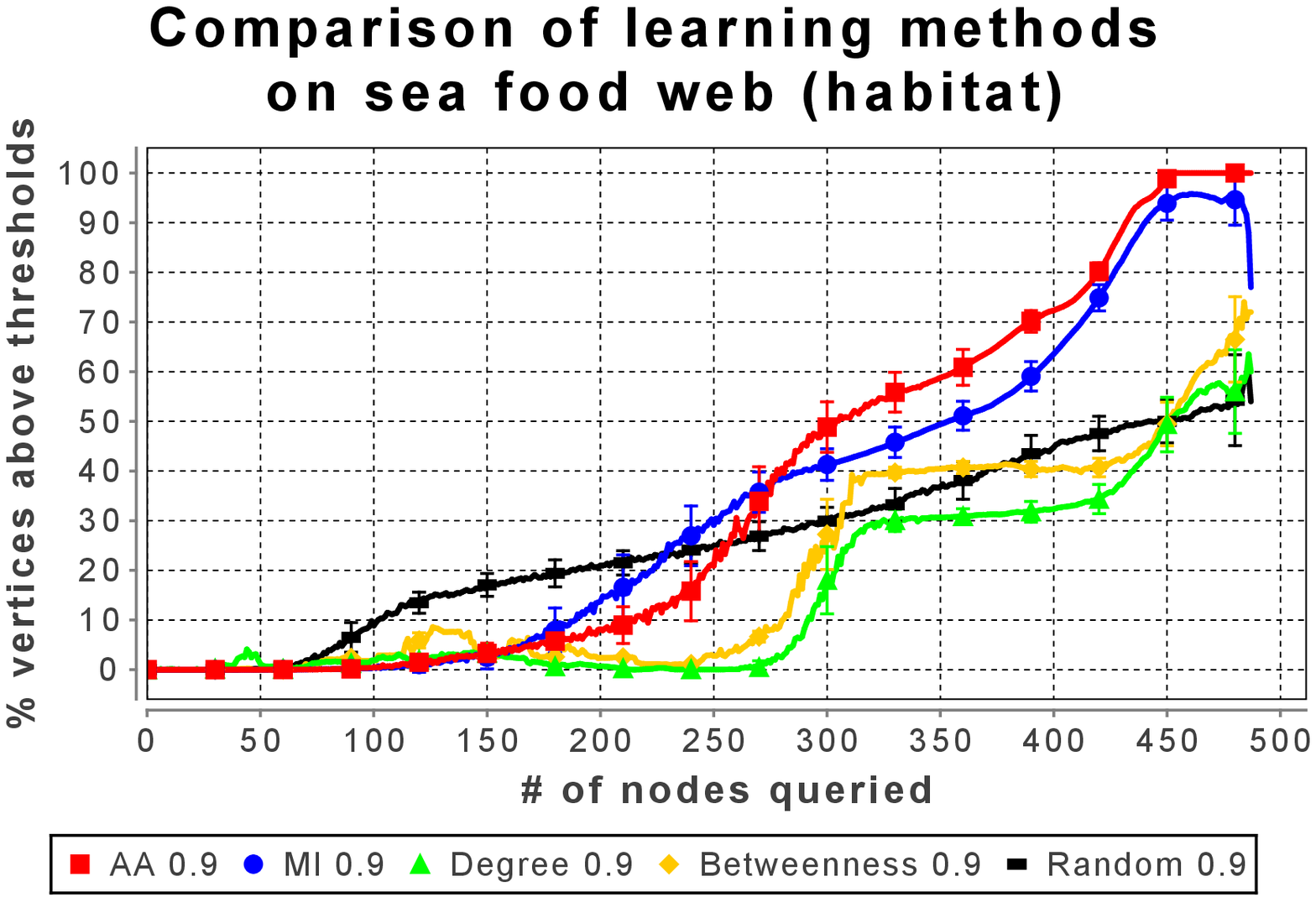}
\includegraphics[width=0.46\textwidth]{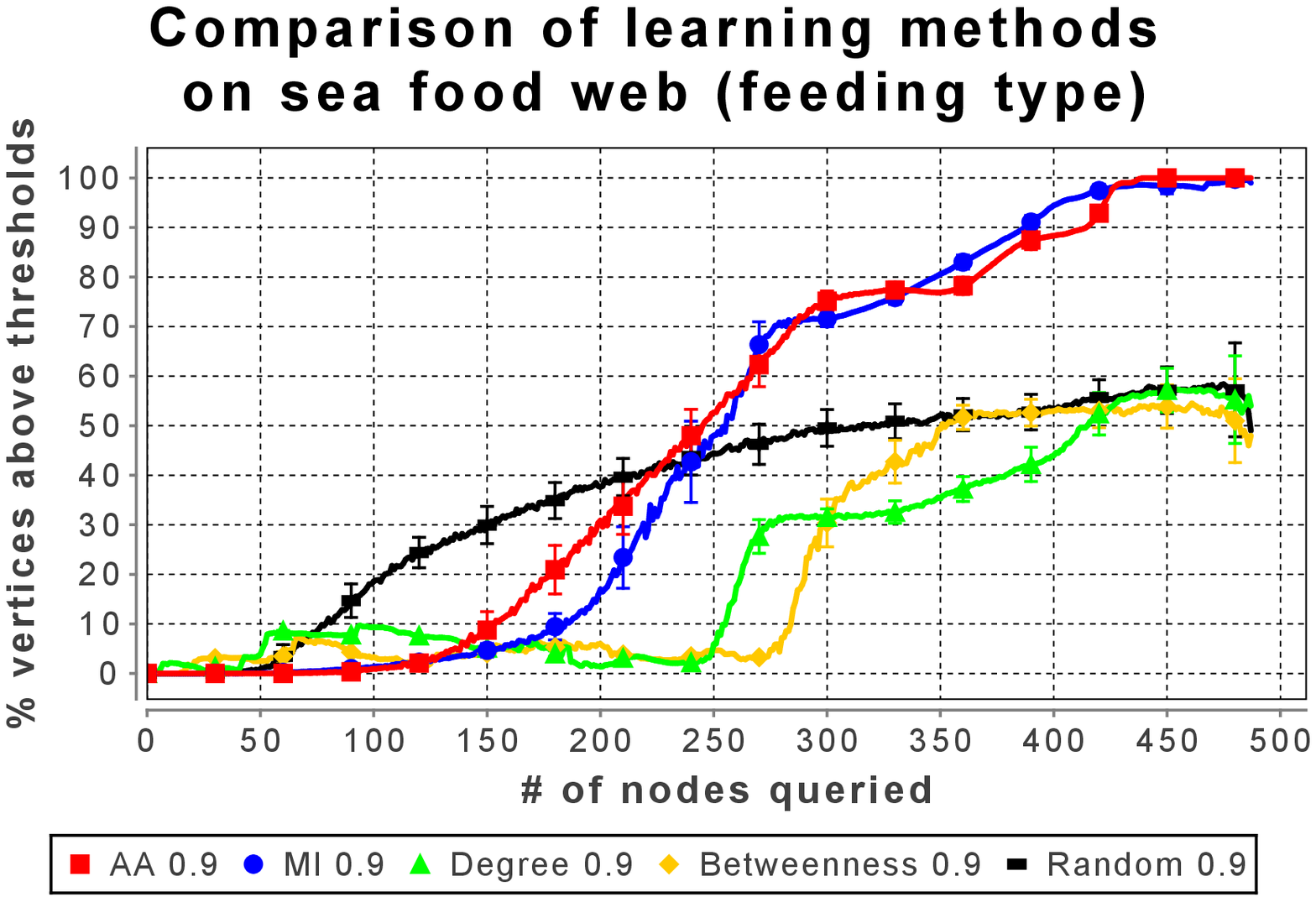}
\caption{A comparison of the MI and AA learning algorithms with three simple heuristics.}
\label{fig:compare}
\end{figure*}

We compared our active learning algorithms with several simple heuristics. These include exploring the node with the highest degree in the subgraph of unexplored nodes, exploring the node with the highest betweenness centrality (the fraction of shortest paths that go through it, see~\cite{brandes,newman_scicolnetwork,newman_betweenness}) in the subgraph of unexplored nodes, and exploring a node chosen uniformly at random from the unexplored ones.
We judge the performance of these heuristics using the same Gibbs sampling process as for MI and AA.

In Fig.~\ref{fig:compare}, we show the results of these heuristics at the $0.9$ accuracy threshold on all three networks, including both the habitat and feeding type variables in the food web.
On Zachary's Karate Club (left) our algorithms outperform these heuristics consistently.  In the \emph{David Copperfield} network (right), the highest-degree and highest-betweenness heuristics enjoy an early lead, but quickly hit a ceiling and are surpassed by MI and AA.


For the Weddell Sea food web (bottom), the highest-degree and highest-betweenness heuristics perform poorly throughout the learning process. One reason for this is that many nodes with high degree or high betweenness are easy to classify from the labels of their neighbors.  By exploring these nodes first, these heuristics leave themselves mainly with hard-to-classify nodes.  The random-node heuristic performs surprisingly well early on, but all three heuristics are worse than MI or AA once they have explored half the nodes.

\section{Conclusion}

Active learning, using mutual information or average agreement coupled with a generative model, offers a new approach to analyzing networks where the topology is known, but knowledge of class labels is incomplete and costly to obtain.  We have shown for three networks, one social, one lexical, and one biological, that our algorithms do a good job of predicting the labels of unexplored nodes after exploring a relatively small fraction of the network, correctly recognizing both assortative and disassortative functional communities.
Certainly not all networks are well-described by the simple block model we use here, but our approach can be generalized to probabilistic network models which take information on the nodes' locations or degrees into account.








\paragraph*{Acknowledgments}
We are grateful to Joel Bader, Aaron Clauset, Jennifer Dunne, Nathan Eagle, Brian Karrer, Jon Kleinberg, Mark Newman, Cosma Shalizi, and Jerry Zhu for helpful conversations, and to Ute Jacob for the Weddell Sea food web data.  J.-B. R. is also grateful to the Santa Fe Institute for their hospitality.  This work was supported by the McDonnell Foundation.



\end{document}